\documentstyle[aps,multicol,epsf]{revtex} \def\tr{{\rm tr}}
\def\CD{{\cal D}}
\def\cl{{\ell}}
\begin{document}
\title{Matrix Generalizations of Some Dynamic Field
Theories}
\author{Mehran Kardar$^{1}$ and A. Zee$^{2}$}
\address{$^1$ Department of Physics,
Massachusetts Institute of Technology,
Cambridge, Massachusetts 02139}
\address{$^2$ Institute for Theoretical Physics, University
of California, Santa Barbara, California 93106-4030 }
\date{\today}
\maketitle
\begin{abstract}
We introduce matrix generalizations of the Navier--Stokes
(NS) equation for fluid flow, and the
Kardar--Parisi--Zhang (KPZ) equation for interface
growth. The underlying field, velocity for the NS equation,
or the height in the case of KPZ, is promoted to a matrix
that transforms as the adjoint representation of $SU(N)$.
Perturbative expansions simplify in the $N\to\infty$ limit,
dominated by planar graphs. We provide the results of a
one--loop analysis, but have not succeeded in finding
the full solution of the theory in this limit.
\end{abstract}
\pacs{}
%%
%\begin{multicols}{2}
%%
\section{Introduction}\label{sIntro}
The occurrence of complex probability distributions under
non-equilibrium conditions is a challenging topic. Recently,
a great deal of effort has been focussed   on the study of
the Kardar--Parisi--Zhang (KPZ) equation\cite{KPZ},
\begin{equation}\label{KPZ}
{\partial h \over \partial t} = \nu \nabla^2 h + {\lambda
\over 2} (\nabla h)^2+\eta( {\bf x}, t),
\end{equation}
a deceptively simple prototype of non-equilibrium
dynamics. The evolution of the field $h({\bf x},t)$, denoting,
for example, the height of a growing interface\cite{reviews},
has a {\it stochastic} component governed by the noise
$\eta( {\bf x}, t)$, usually assumed to be Gaussian distributed, with
zero mean, and correlations
\begin{equation}\label{noise}
\left\langle\eta({\bf x}, t) \eta( {\bf x'}, t')\right\rangle= 2D
\delta^d({\bf x-x'}) \delta(t-t').
\end{equation}
The dynamical equation can be regarded as representing an
elaborate filter that converts the simple correlations of the
noise $\eta$ to complex correlations of the field $h( {\bf x}, t)$.
For example, the two point
correlations satisfy the dynamic scaling form,
\begin{equation}\label{defhh}
\left\langle \left[ h({\bf x},t)- h({\bf
x'},t')\right]^2\right\rangle=
|{\bf x-x'}|^{2\chi} f\left(|{\bf x-x'}|^z/|t-t'| \right),
\end{equation}
where $z$ and $\chi$ are the so called dynamic and
roughness exponents.

For $\lambda=0$, the linear diffusion equation gives $z=2$
and $\chi=(2-d)/2$. Equation (\ref{KPZ}) is rendered
interesting and non-trivial due to
the nonlinearity. The renormalization group (RG) flow of the
effective coupling constant
$g^2 \equiv \lambda^2 D/\nu^3$, is given in a one loop
perturbative calculation\cite{KPZ,FNS} as
\begin{equation}\label{RGg1}
{dg^2\over d\cl}=g^2\left[2-d +{g^2\over 2}K_d {2d-3\over
d}\right],
\end{equation}
where $K_d=S_d/(2\pi)^d$, and $S_d$ is the $d$-dimensional
solid angle. Equation (\ref{RGg1}) suggests the presence of
strong coupling behavior in all dimensions $d$. In $d=1$
the dynamics is super-diffusive, characterized by
exponents $z=3/2$ and $\chi=1/2$\cite{FNS}. At $d=2$,
the nonlinearity is marginally relevant, signaling strong
coupling behavior; recent numerical
studies\cite{Ala-Nissila} indicate $z \approx 1.612$, and
$\chi \approx 0.386$. In dimensions $d>2$, a critical
strength of $\lambda$ separates diffusive [$z=2$ and
$\chi=(2-d)/2$], and non-trivial regimes. Unfortunately,
the strong coupling fixed point is not accessible by
perturbative RG\cite{FNS,Medina}, which has recently
been extended to two loop order\cite{Sun,Frey}. In an
alternative approach, the one loop perturbation
equations are converted into a set of self-consistent,
so-called mode-coupling
equations\cite{Schwartz,Bouchaud,Doherty,Tu}. The
numerical solution of these equations remains
controversial, with glassy solutions recently suggested as
a possibility\cite{Claudin}. The mode-coupling equations
have been shown to be exact when the field $h$ is
generalized to an $N$ component {\it vector},
in a large $N$ limit.

The KPZ equation bears certain similarities (but also
profound differences) with the Navier--Stokes (NS)
equation for fluid flow. The fluid velocity
field ${\bf v}({\bf x},t)$ evolves according to
\begin{equation}\label{NS}
{\partial {\bf v} \over \partial t} + \lambda ({\bf v}\cdot
{\bf\nabla}){\bf v}= -{{\bf \nabla}p \over \rho}+\nu
\nabla^2 {\bf v} +{\bf f}( {\bf x}, t),
\end{equation}
where $\nu$ is the viscosity, $\rho$ is the density, and
$\lambda$ (=1) is introduced for book-keeping purposes.
The pressure $p( {\bf x}, t)$ is adjusted to enforce the
incompressibility condition ${\bf\nabla}\cdot {\bf v}=0$.
The qualitative difference with the KPZ equation is in the
stirring force ${\bf f}( {\bf x}, t)$, which has non-zero
Fourier components ${\bf f}( {\bf k}, \omega)$ only at
small $k\sim 1/L$, where $L$ is the length scale at which
energy is pumped into the system. The nonlinearity
then transfers the energy into modes with higher
wave-numbers $k$, eventually dissipating it at the
dissipative length scale. This establishes the
{\it Kolmogorov energy cascade} in the intermediate
(inertial) regime of wave-vectors\cite{Kolmogorov}.
A basic question in turbulence is the nature of this
energy cascade, usually described by a scaling form
for the energy spectrum, $E(k)\propto k^{-\zeta}$. (If the
velocity correlations satisfy a dynamic scaling form as
in Eq.(\ref{defhh}), then $$E \equiv \int_0^{\infty} dk E(k)
\sim \int d^dk d\omega \int d^dx dt e^{i({\bf k}.{\bf x} - \omega t)}
\left\langle {\bf v}({\bf x},t){\bf v}({\bf
0},0)\right\rangle$$ and thus $$E(k) \sim k^{d-1} \int d^dx e^{i{\bf
k}.{\bf x} } x^{2\chi}$$ and thus, as is well known, we have the exponenet
relation $\zeta=2\chi+1$.) The famous
argument by Kolmogorov\cite{Kolmogorov} gives
$\zeta=5/3$, in reasonable agreement with experiment.

To apply the methods of dynamical RG\cite{FNS,deDM}, it is
usually assumed that ${\bf f}( {\bf k}, \omega)$ is Gaussian
distributed with zero mean, and correlations
\begin{equation}\label{NSnoise}
\left\langle f_i({\bf k}, \omega) f_j({\bf k'},
\omega')\right\rangle=
2D(k)P_{ij}({\bf k}) (2\pi)^{d+1} \delta^d({\bf k}+{\bf k'})
\delta(\omega+\omega').
\end{equation}
The transverse projection operator $P_{ij}({\bf k})\equiv
\delta_{ij}-k_ik_j/k^2$ is a consequence of the condition
${\bf \nabla}\cdot{\bf f}=0$. Thermal fluctuations in the
fluid can be modelled by $D(k)\propto k^{2}$, and lead
to the familiar long-time tails in the velocity correlation
functions\cite{FNS}. More drastic forms of stirring are
modelled by $D(k)\propto  k^{-p}$ with $-p<2$\cite{deDM}.
In the problem of turbulence, the stirring at the longest
scales resembles $D(k)\propto \delta^d ({\bf k})$, which has
the same scaling as $p=d$. This observation motivates
some of the more recent applications of RG to
turbulence\cite{YO}. An extensive review of the
application of RG to turbulence in given by Mou and
Weichman in Ref.\cite{MW}. In particular, these authors
develop a set of self-consistent equations for the problem,
exact in an appropriate $N\to\infty$ limit, for which
$\zeta=3/2$.

The study of large $N$ matrix theories started with the
classic work of Wigner\cite{WIG} and has been
developed\cite{POR,MEH} over the years. Notable
advances include the formulation of large $N$ quantum
chromodynamics\cite{thooft}, and applications to random
surfaces\cite{jerusalem}. Intensive studies of the subject
have continued in recent years\cite{bz}. Here we
formulate and study large $N$ {\it matrix}
generalizations of the KPZ and NS equations by taking
advantage of  a crucial difference between vector  and
matrix models: the product of two vectors is not a vector
while the product of two matrices is a matrix. Thus, we
are able to promote the fields $h({\bf x},t)$ and
${\bf v}({\bf x},t)$ to $N\times N$ hermitian matrices and
preserve the non-linear structure of the $N=1$
equations. As large $N$ matrix models are generally
quite different from their vector counterparts, we may
hope to obtain results that are distinct from the earlier
mode-coupling equations\cite{Doherty}. In particular,
it is possible that techniques developed in large $N$
matrix theory, such as the fact that its perturbative
expansion is dominated by planar diagrams, may be
brought to bear on this problem, leading to new
analytic results. Although the primary interest is in the
$N=1$ limit, analytical insights may shed light on
such controversial issues as the existence or
absence of an upper critical dimension for the KPZ
equation. The main focus of this paper is a perturbative
analysis of the generalized KPZ equation introduced in
Sec.\ref{sKPZ}. A brief analysis of the matrix NS
equation is presented next in Sec.\ref{sNS}.
Prospects for further developments are discussed in
the final Section, \ref{sDis}.

\section{Matrix KPZ equation}\label{sKPZ}

We promote the height $h({\bf x}, t)$ in Eq.(\ref{KPZ}) to a
hermitian matrix field $h^\alpha_\beta({\bf x}, t)$,
transforming as the adjoint representation
under the $SU(N)$ symmetry group. The indices $\alpha$
and $\beta$, which we call color indices, run over
$1, \cdots,N$, with $N$ large. Time evolution of the matrix
is governed by the generalized KPZ equation,
\begin{equation}\label{nKPZgeneral}
{\partial h^\alpha_\beta \over \partial t} =\nu_1 \nabla^2
h^\alpha_\beta
+{\nu_2}\delta^\alpha_\beta \nabla^2 h^\gamma_\gamma +
{\lambda_1 \over 2}
\nabla h^\alpha_\gamma \nabla h^\gamma_\beta
+{\lambda_2} \nabla
h^\alpha_\beta \nabla h^\gamma_\gamma + {{\lambda_3
\over 2}
\delta^\alpha_\beta \nabla h^\gamma_{\gamma'} \nabla
h^{\gamma'}_\gamma
+{\lambda_4 \over 2} \delta^\alpha_\beta \nabla
h^\gamma_\gamma \nabla
h^{\gamma'}_{\gamma'}} +\eta^\alpha_\beta( {\bf x}, t).
\end{equation}
All repeated indices are summed over; the contractions
guarantee the $SU(N)$ invariance. The stochastic noise
is assumed to have zero mean, with correlations
\begin{equation}\label{nnoisegeneral}
\left\langle\eta^\alpha_\beta({\bf x},
t)\eta^{\beta'}_{\alpha'}({\bf x'},
t')\right\rangle
=2\left[D_1 \delta^\alpha_{\alpha'} \delta^{\beta'}_\beta + D_2
\delta^\alpha_\beta \delta^{\beta'}_{\alpha'}\right]\delta^d
({\bf x-x'})\delta(t-t').
\end{equation}
Although constrained by $SU(N)$ invariance, the equation
still admits 8 parameters; four at the linear level
($\nu_{1,2}$, $D_{1,2}$) and four nonlinearities
($\lambda_{1,2,3,4}$). Due to this large number of
parameters, Eqs.(\ref{nKPZgeneral}) are quite formidable. It
would be convenient if we could focus on a subset of
parameters such as $(\nu_1,\lambda_1,D_1)$. To check
if such a simplification is possible, we appeal to a graphical
perturbation expansion. The diagrammatic
approach\cite{Medina} to the KPZ equation may be
generalized to the matrix $h^\alpha_\beta$ represented
by two points: The propagator is a double line that keeps
track of the flow of the color indices $\alpha$, $\beta$, as
well as momenta and frequencies. The nonlinearities are
represented as triplets of double lines, while averaging
over noise joins pairs of such lines. These diagrammatic
entities are depicted in Fig.~(\ref{F1}). By substituting
these entities for the corresponding elements, each term
in the original perturbative expansion now has several
counterparts in the matrix theory. Summing over the
{\it internal} labels generates a factor of $N$ for every
closed loop. Keeping only the diagrams with the largest
power of $N$ will hopefully lead to some simplification.

To gain insight into the structure of the perturbation series,
we started by setting all parameters except
$(\nu_1,\lambda_1,D_1)$ to zero, and performing a
{\it one loop} RG analysis. From the perturbative expansion
of the propagator, and the correlation function, we obtain
respectively
\begin{eqnarray}\label{RGone}
{d\nu_1\over d\cl}=\left(z-2\right)\nu_1-{K_d(d-2)\over
4d}\cdot {N\lambda_1^2D_1\over \nu_1^2}&\qquad &
{d\nu_2\over d\cl}=\left(z-2\right)\nu_2+{K_d(d-2)\over
4d}\cdot {\lambda_1^2D_1\over \nu_1^2},\nonumber \\
{dD_1\over d\cl}=\left(z-2\chi-d\right)D_1+{K_d\over
4}\cdot {N\lambda_1^2D_1^2\over \nu_1^3} &\qquad&
{dD_2\over d\cl}=\left(z-2\chi-d\right)D_2+{K_d\over 4}
\cdot{\lambda_1^2D_1^2\over \nu_1^3} .
\end{eqnarray}
%($K_d$ is a constant related to the $d$--dimensional solid angle.)
The equations for $\nu_1$ and $D_1$ are similar to those of the
scalar KPZ, with $2\lambda^2$ replaced by $N\lambda_1^2$.
We also see that $\nu_2$ and $D_2$ are already generated at
this order, although these corrections are smaller by a factor of
$1/N$. The reductions by powers of $1/N$ arise because to
create the pairing of indices (the `U' turns) corresponding to
$\nu_2$, etc., out of $(\nu_1,\lambda_1,D_1)$, the lines have
to cross at some point. However, as is well known in matrix theory,
such {\it non--planar} diagrams carry smaller powers of $N$, and
can be neglected in the $N\to\infty$ limit\cite{thooft}. The next
question is whether these generated parameters, albeit small,
can feed back into the recursion relations for
$(\nu_1,\lambda_1,D_1)$ significantly, so that their inclusion
is necessary. Clearly, replacing any of the original graphical
elements with the above generated ones results in a
smaller power of $N$ from their smaller magnitude. The
question is whether the rearrangements of the indices
can generate compensating powers of $N$. For all
simple graphs that we examined the answer was negative.

We thus assume that the subspace $(\lambda_1\equiv
\lambda/\sqrt{N}, \nu_1\equiv\nu,D_1\equiv D)$ is closed
up to order $1/N$, and study the matrix equation
\begin{equation}\label{nKPZ}
{\partial h^\alpha_\beta \over \partial t} = \nu \nabla^2
h^\alpha_\beta +
{\lambda \over 2\sqrt{N}} \nabla h^\alpha_\gamma \nabla
h^\gamma_\beta +\eta^\alpha_\beta( {\bf x}, t),
\end{equation}
with a noise term described by
\begin{equation}\label{nnoise}
\left\langle \eta^\alpha_\beta({\bf x},
t)\eta^{\beta'}_{\alpha'}({\bf x'}, t')\right\rangle
=2D \delta^\alpha_{\alpha'} \delta^{\beta'}_\beta
\delta^d ({\bf x-x'}) \delta(t-t').
\end{equation}
(We shall shortly demonstrate that this choice of scaling the
parameters with $N$, which is in line with the one loop result,
leads to a consistent large $N$ theory at all orders.)
The reduced matrix equation is invariant under the
transformation
\begin{equation}\label{GI}
h'^\alpha_\beta({\bf x},t)=h^\alpha_\beta\left({\bf
x}+{\lambda\over \sqrt{N}}{\bf u}t,t\right)
+\delta^\alpha_\beta {\bf u\cdot x}+
\delta^\alpha_\beta{\lambda\over 2\sqrt{N}} u^2t,
\end{equation}
which generalizes the so called Galilean invariance\cite{FNS}
of the scalar KPZ equation. Referring to Eq.(\ref{defhh}),
we see that the dynamic and roughness exponents are defined
by requiring that the effective equation preserves its form under
the transformation ${\bf x} \rightarrow b {\bf x}$,
$t \rightarrow b^z t$, and $h \rightarrow b^{\chi} h$. Under
these rescalings, and with $b=e^\ell$, the non-linear
coupling $\lambda$ evolves according to
\begin{equation}\label{lambda}
{d\lambda\over d\cl}=\left( \chi+z-2 \right)\lambda, \quad
\end{equation}
quite generally for any $N$. To preserve the invariance
in Eq.(\ref{GI}) under the above rescalings, the
coefficient $\lambda$ must remain constant, leading to the
exponent relation
\begin{equation}\label{GIexponents}
\chi+z=2.
\end{equation}
This identity holds at any fixed point with finite $\lambda$.

In $d=1$, Eq.(\ref{nKPZ}) also satisfies a fluctuation--dissipation
condition\cite{FDT} which implies that the probability
distribution
\begin{equation}\label{SS}
{\cal P}_0[h]\propto\exp\left[ -{\nu\over 2D}\int dx\tr\left(
\partial_x h \right)^2 \right],
\end{equation}
is a stationary solution of the Fokker--Planck equation for
${\cal P}([h],t)$\cite{ITU}.  In deriving the matrix version of
the Fokker--Planck equation we must be careful about the
non-commutativity of matrices. Going through the usual
formal steps, we obtain
\begin{equation}\label{matrixfp}
{\partial\over\partial t}{\cal P}[h,t]
= -\int d^dx \left\{\tr {\delta\over\delta h}\left[\left(
\nu \nabla^2 h+ {\lambda\over 2\sqrt N} {(\nabla h)}^2
\right) {\cal P}\right]-D\tr \left({\delta^2\over
\delta h^2}{\cal P}\right)\right\}
\end{equation}
The nesting of the parentheses in the first term of this equation requires
some explanation: the large round brackets contain  all the matrix
quantities to be traced over (${\cal P}$ is of course not a matrix), while
the square parenthesis indicates that ${\delta/\delta h}$ should also
act on ${\cal P}$. Inserting Eq.(\ref{SS}), we find that the right hand
side of Eq.(\ref{RGone}) becomes,  for a general dimension $d$,
proportional to $\int d^dx\, \tr[{(\nabla h)}^2 {\nabla}^2 h].$ For $d=1$
the integrand is a total divergence, indicating that ${\cal P}_0$ is a
stationary solution. For this stationary state $\chi=1/2$ (and $z=3/2$
from Eq.(\ref{GIexponents})), also in agreement with the RG results of
Eq.(\ref{RGone}). Thus, in $d=1$ the exponents for $N=1$ and
$N\to\infty$ are identical; also a feature of the vector
equations\cite{Doherty}. In this context, it is interesting to note a
recent study of a trimer deposition model on a line that is claimed
to be equivalent to an $N=2$ matrix model\cite{Barma}.
Numerical results on this model appear to suggest
$z\approx 2.5$.

By integrating over the noise in a functional integral
description, we can also obtain a more traditional field
theoretical formulation of the problem as
\begin{equation}\label{field}
\left\langle Z \right\rangle=\int \CD h \int \CD\eta \, J[h]
\delta \left({\partial h^\alpha_\beta \over \partial t} -
\nu \nabla^2 h^\alpha_\beta
- {\lambda \over 2\sqrt{N}} \nabla h^\alpha_\gamma \nabla
h^\gamma_\beta -\eta^\alpha_\beta( {\bf x}, t)\right)
\exp\left[-{1\over 2D}\tr\,\eta^2\right]\equiv \int \CD h\,
e^{-S(h)},
\end{equation}
where $J[h]$ is the Jacobian associated with the
transformation from $h$ to $\eta$. The value of $J[h]$
depends on the discretization of the evolution
equation. In an Ito choice of discretization in which the noise
$\eta({\bf x},t)$ only affects the field at $h({\bf x},t+\tau)$,
the Jacobian associated with
$|\partial \eta({\bf x},t)/\partial h({\bf x},t+\tau)|$ is simply a
constant\cite{Ito}. We shall thus ignore this factor henceforth.
The $\delta$--functions in Eq.(\ref{field}) can be implemented
by using a conjugate field $\tilde{h}$ as in Ref.\cite{Frey}.
Integrating over this field, as well as the noise $\eta$,
then leads to the action
\begin{equation}\label{action}
S(h) ={1\over 2D}\int d^d{\bf x}\, dt \,\tr \left({\partial h
\over
\partial t} -\nu \nabla^2 h -{\lambda \over 2\sqrt{N}}
(\nabla h)^2\right)^2.
\end{equation}
We can now remove two of the remaining parameters by
the rescalings, $t \rightarrow t/\nu$ and
$h \rightarrow \sqrt{ND/\nu} \,h$, leading to
\begin{equation}\label{action2}
S(h) = {N\over 2}\int d^d{\bf x}\, dt \, \tr \left( \left({\partial
\over\partial t} - \nabla^2\right) h
-{g \over 2 } (\nabla h)^2\right)^2,
\end{equation}
with $g$ as defined after Eq.(\ref{RGg1}). With its cubic
and quartic interactions, the above action is
reminiscent of that of a non-abelian gauge theory such
as quantum electrodynamics without quarks.

The Feynman rules corresponding to Eq.(\ref{action2}) are
given in Fig.(\ref{F2}). Each propagator line has a factor of
$1/N$, while each vertex is proportional to $N$, and each
loop generates a factor of $N$. Thus, a given diagram with
$E$ external lines, $I$ internal lines, $V_3$ three-point
vertices, $V_4$ four-point vertices, and $L$ loops, is
associated with a factor of $N^P$ with $P = L+V_3 + V_4-I$.
Using the standard topological identity
$L = I -V_3 - V_4 +1$, we obtain $P=1$. For $E = 3$ and
$E = 4$; this certainly coincides with the bare scalings of
three and four point vertices in Eq.(\ref{action2}). Thus the
choice of scaling the nonlinearity with $\sqrt{N}$ indeed
leads to a consistent large $N$ expansion.

We also carried out a one--loop perturbative RG, directly on
the action in Eq.(\ref{action2}). The basic idea is to integrate
out short wavelength modes, with $\Lambda/b<k<\Lambda$
where $\Lambda$ is a cutoff. The effective action for the
remaining modes then has a leading gradient
expansion of the form
\begin{equation}\label{action3}
\tilde S(h) = {N\over 2}\int d^d{\bf x}\, dt \, \tr \left(
\left(\alpha{\partial \over \partial t} - \beta \nabla^2\right)
h -\alpha{g\over 2 } (\nabla h)^2 \right)^2,
\end{equation}
depending on two parameters $\alpha$ and $\beta$.
The same coefficient $\alpha$ multiplies both
$\partial_t h$ and the nonlinearity, as required by the
Galilean symmetry in Eq.(\ref{GI}). To calculate the
effective action, we only have to look at terms quadratic
in $h$. Thus, it suffices to evaluate the renormalized
propagator.

Note that a term proportional to $\tr (\nabla h)^2$ is
also allowed by symmetries, and is in fact generated.
However, such a term can be eliminated by
transforming to a moving coordinate frame
$h\to h+ct$. Thus, in calculating the
one-particle-irreducible self-energy function
$\Sigma(k, \omega)$, we only have to extract the
coefficients of the $\omega^2$ and $k^4$ terms in
a low frequency and wave-number expansion. The
one--loop corrections to the propagator are indicated in
Fig.~(\ref{F3}). In fact, the two diagrams are identical,
except that one is smaller by a factor of $1/N$ due to
the crossing of lines. We shall keep track of both
diagrams by including a factor of $(1+1/N)$, thereby also
correctly reproducing the $N=1$ case\cite{noteN}.
A simplifying feature is that  the quartic interaction term
$\sim g^2 (\nabla h)^4  $ in Eq.(\ref{action2}) does not
enter into the calculation to this order. Evaluating the
frequency and momentum dependence of the
remaining self--energy diagram is cumbersome,
but straightforward, and leads to
\begin{equation}\label{Sigma}
\Sigma(k,\omega)={1\over 4}\left[\omega^2+{4d^2-d-
6\over d(d+2)}k^4\right] \int_{\Lambda/b}^\Lambda
{d^d q\over (2\pi)^d}\cdot{1\over q^2}.
\end{equation}
Adding the self-energy to the bare propagator, we obtain
the renormalization parameters,
\begin{eqnarray}\label{alphabeta}
\alpha&=&1-{g^2\over 16}\left( 1+{1\over N} \right)K_d
d\cl ,\\
\beta&=&1-{g^2\over 16}\left( 1+{1\over N} \right){4d^2-
d-6\over d(d+2)}K_dd\cl .
\end{eqnarray}
We have evaluated the integral in Eq.(\ref{Sigma}) over an
infinitesimal shell by setting $b=1+d \cl$ and writing
$\int_{\Lambda/b}^\Lambda {d^d q\over (2\pi)^d}\cdot
{1\over q^2}=K_d d \cl$, with
$K_d=S_d\Lambda^{d-2}/(2\pi)^d$,
where $S_d$ is the $d$-dimensional solid angle.

The cutoff in the effective action can be restored to its
original value by setting ${\bf x}=b{\bf x'}$,
accompanied by $t=b^z t'$ and $h=b^\chi h'$.
The parameters in the quadratic part of the renormalized
action can be reset to unity by choosing the exponents
\begin{eqnarray}\label{zchi1}
z&=&2+{g^2\over 16}K_d\left( 1+{1\over N} \right){3(d-
2)(d+1) \over
d(d+2)},\\ \chi&=&{2-d\over2}+{g^2\over 32}K_d\left(
1+{1\over N}\right){(d-1)(5d+6)\over d(d+2)}.
\end{eqnarray}
Finally, the flow of the coupling constant is given by
$dg/d\cl=\left( \chi+z-2 \right)g$, resulting in
\begin{equation}\label{RGg2}
{dg^2\over d\cl}=g^2\left[2-d +{g^2\over 16}K_d \left(
1+{1\over N} \right)
{11d^2-5d-18\over d(d+2)}\right].
\end{equation}

Surprisingly, this flow equation looks quite different from
that obtained by standard dynamic RG
methods\cite{Medina}, which generalize Eq.(\ref{RGg1}) to
\begin{equation}\label{RGg3}
{dg^2\over d\cl}=g^2\left[2-d +{g^2\over 4}K_d \left(
1+{1\over N} \right) {2d-3\over d}\right].
\end{equation}
The resolution to this discrepancy may be due to the fact
that the flow equations are not fundamental. The two
equations in fact agree for both $d=2$ and $d=1$.
In $d=1$, the RG is constrained to give the correct
exponents due to the symmetries embodied by
Eqs.(\ref{GI}) and (\ref{SS}). It is indeed easy to check
that by substituting $K_1(1+1/N)g^{*2}=4$ in
Eqs.(\ref{zchi1}) we recover $z=3/2$ and $\chi=1/2$.
The exponents at the phase transition between the
weak and strong coupling phases in $d=2+\epsilon$
dimensions are also correctly given by
$z=2+O(\epsilon^2)$, $\chi=0+O(\epsilon^2)$. (The
correlation length at this transition diverges with an
exponent $\nu=\epsilon^{-1}+O(1)$.) Thus all physical
quantities calculated from Eqs.(\ref{RGg2}) and
(\ref{RGg3}) appear to be identical. The coefficient
of the quadratic term in Eq.(\ref{RGg2}) changes sign at
$d=1.5265$ as opposed to $d=3/2$ for Eq.(\ref{RGg3}).
However, there is probably no physical significance to
this dimension, as it disappears in the two loop
calculation of Ref.\cite{Frey}. Even simpler flow equations
(without any higher order terms) are obtained in an RG
of the KPZ equation that proceeds
from a mapping to directed polymers\cite{Lassig}.

\section{Matrix NS equation}\label{sNS}
For the Navier-Stokes equation we promote the velocity
vector field ${\bf v}=(v_1,\cdots,v_d)$ to a set of
$N\times N$ hermitian matrices
${\bf v}^\alpha_\beta({\bf x}, t)$, imposing the
incompressibility $\partial_i v_i=0$ as before. The
non-linear term has four matrix counterparts
$v_j(\partial_j v_i)$, $(\partial_j v_i) v_j$,
$v_j(\partial_i v_j)$, and $(\partial_i v_j) v_j$. We will
suppress the matrix indices henceforth.
(We shall neglect various linear and nonlinear terms
that can be constructed using combinations of the unit
matrix and/or $\tr {\bf v}$ as in Eq.(\ref{nKPZgeneral}).
They are again expected to be irrelevant in the large
$N$ limit.) We further require the matrix generalization
of Eq.(\ref{NS}) to satisfy two important physical
conditions: (1) a suitable generalization of
Galilean invariance, and (2) conservation of energy in the
absence of dissipation. Subject to the above limitations,
we arrive at
\begin{equation}\label{matrixNS}
{\partial v_i \over \partial t} + {\lambda\over 2\sqrt{N}}
\left\{v_j, \partial_j v_i\right\}+{\kappa\over 2\sqrt{N}}
\left[v_j,\partial_j v_i+\partial_i v_j\right]
=-{\partial _i p \over \rho}+
\nu \partial_j^2 v_i +f_i( {\bf x}, t),
\end{equation}
where the curly ($\left\{  ... \right\}$) and square ($[...]$)
brackets indicate, respectively, matrix
anti-commutation and commutation.
(The matrix products ensure $SU(N)$ invariance.)
The commutator term proportional to $\kappa$ has
no scalar counterpart. The `pressure' matrix
$p( {\bf x}, t)$, is again adjusted to enforce the
incompressibility condition $\partial_i v_i=0$.
Finally, it is assumed that the random (matrix) force
is Gaussian distributed with zero mean, and
correlations that in Fourier space are given by
\begin{equation}\label{NSnoise}
\left\langle \left(f^\alpha_\beta\right)_i({\bf k}, \omega)
\left(f^{\beta'}_{\alpha'}\right)_j({\bf k'},
\omega')\right\rangle=2D(k)\delta^\alpha_{\alpha'}
\delta^{\beta'}_\beta P_{ij}({\bf k}) (2\pi)^{d+1}
\delta^d({\bf k}+{\bf k'}) \delta(\omega+\omega').
\end{equation}
The transverse projection operator  again follows from
incompressibility, and $\partial_i f_i=0$.

A form of Galilean invariance is satisfied by the generalized
equation, if in a frame of reference moving with velocity
${\bf u}$, the generalized velocity transforms to
\begin{equation}\label{NSmatrixGI}
{\bf v'}^\alpha_\beta({\bf x},t)={\bf v}^\alpha_\beta\left({\bf
x}+{\lambda\over \sqrt{N}}{\bf u}t,t\right)
+{\bf u}\delta^\alpha_\beta .
\end{equation}
It is easy to check that Eq.(\ref{matrixNS}) is invariant
under this transformation. A natural choice for the
generalized energy of the fluid is $E={\rho/2}\int
d^dx \tr\left(v_i v_i\right)$. In the absence of dissipation
($\nu=0$) and forcing (${\bf f}={\bf 0}$), the change
in energy is governed by
\begin{equation}\label{NSenergy1}
{dE \over dt}=\rho\int d^dx\tr\left(v_i \partial_t v_i\right)=
-\int d^dx\tr\left({\lambda\rho \over 2\sqrt{N}}v_i\left\{
v_j,\partial_j v_i\right\}+{\kappa\rho \over 2\sqrt{N}}
v_i\left[v_j, \partial_j v_i+\partial_i v_j\right]+
v_i\partial_ip\right).
\end{equation}
Using the incompressibility condition $\partial_j v_j=0$, and
reordering the matrices inside the trace, we can transform
the above expression to
\begin{equation}\label{NSenergy2}
{dE \over dt}=-\int d^dx\partial_j \tr\left[{\lambda\rho
\over 2\sqrt{N}}v_i v_iv_j +v_j p\right]=0,
\end{equation}
i.e. energy is conserved as required. In what follows we will set $\kappa$
to zero for simplicity. It would be interesting to study how $\kappa$ flows
under the renormalization group.

The above conditions are sufficient to give the scaling
exponents for Eq.(\ref{matrixNS}) under most forcing conditions.
Just as before, we define the exponents $\chi$ and $z$
through the correlation function
\begin{equation}\label{nscorre}
\left\langle v_i({\bf x},t)v_j({\bf x'},t')\right\rangle=
\delta_{ij}
|{\bf x-x'}|^{2\chi} f\left(|{\bf x-x'}|^z/|t-t'| \right),
\end{equation}
As  in Eq.(\ref{GIexponents}), the requirement of generalized
Galilean invariance leads to the exponent relation $\chi+z=1$.
We need one more exponent relation to determine
$\chi$ and $z$ separately.

Let us first examine thermal noise with $D(k)=Dk^2$ (model A in
the language of Ref.\cite{FNS}). In this case, the linearized
equation with $\lambda=0$ has a steady state probability
distribution
\begin{equation}\label{NSSS}
{\cal P}_0[{\bf v}]\propto\exp\left[ -{\nu\over 2D}\int
d^dx\tr\left( {\bf v}^2 \right) \right].
\end{equation}
It is then straighforward to show that the contribution of the
non-linear term ($\lambda\neq 0$) to the probability
current of the Fokker--Planck equation is the integral of a
divergence. The manipulations establishing this result
are almost identical to those in Eqs.(\ref{NSenergy1}) and
(\ref{NSenergy2}). Thus by ensuring energy conservation,
we have also set up a fluctuation--dissipation
condition\cite{FDT}. From Eq.(\ref{NSSS}) we can read
off the dimension of the velocity as $\chi=-d/2$, leading to
$z=1+d/2$ from Galilean invariance. (Note that in
$d=1$, Eq.(\ref{matrixNS}) for model A is simply the
derivative of Eq.(\ref{nKPZ}).)

Now consider a more general stirring force with
$D(k)=D_pk^{-p}$. Because the nonlinearity is
proportional to the gradient, the perturbative series
generates only powers of $k^2$. Thus the coefficient $D_p$,
the most relevant component of noise as $k\to 0$, is not
renormalized to any order in perturbation theory.
This non-renormalization leads to the exponent
relation\cite{deDM} $\chi=(z-d+p)/2$. Together
with the constraint from Galilean invariance, we
obtain $\chi=(1-d+p)/3$ and $z=(2+d-p)/3$, and a
Kolmogorov exponent of $\zeta=2\chi+1=(5-d+p)/3$.
For the case of large scale stirring ($p=d$), the
Kolmogorov result of $\zeta=5/3$ is recovered.
However, by appealing to this scaling analysis,
we have simply reproduced previously known RG results,
and gained nothing new from the matrix equation.
By contrast, the large $N$ equations introduced by Mou
and Weichman\cite{MW} can be summed exactly in
a large $N$ limit. These authors find that the perturbative
exponents break down for $-p\leq 1-d$, when the
exponent $z$ sticks to 1. The non-linear term now
dominates $\partial_t {\bf v}$, and equating the bare
dimensions of ${\bf v\cdot\nabla v}$ and the noise leads
to $\chi=(1-d+ p)/4$. According to Ref.\cite{MW}, for
$p\geq d$, this leads to the exact result of $\zeta=3/2$
for the Kolmogorov exponent in their large $N$
generalization of the NS equation. Unfortunately, we have
not yet succeeded in finding the exact behavior of our
matrix generalization as $N\to\infty$. Mou and
Weichman\cite{MW} do consider an adjoint $SU(N)$
generalization, presumably equivalent to
Eq.(\ref{matrixNS}). However, we believe that the
construction of the equation, and its analysis, is more
transparent in our matrix formulation.

\section{Discussion}\label{sDis}

In this paper, we introduced and analyzed matrix
generalizations of the KPZ and NS equations.
It is clearly possible to construct similar
generalizations of other dynamical equations, such as for
the time dependent Landau--Ginzburg process. Our
original motivation was to find a set of closed form
expressions that are exact in the large $N$ limit, and
which hopefully yield different results from those based
on other $N\to\infty$ generalizations of these
equations\cite{Doherty,Tu,MW}. The large $N$ limit
does indeed lead to certain simplifications of
perturbative expansions; most notably in the
dominance of planar diagrams\cite{thooft}.
However, we have yet to succeed in taking advantage of this
fact to reduce the perturbation series to a closed set of
self-consistent equations.Nevertheless, we are hopeful that
progress towards this goal can be made. A number of
techniques for dealing with large $N$ matrix theories have
been developed over the years. For instance, consider the
model where a single $N\times N$ hermitian matrix $\varphi$
is taken randomly from the probability ensemble $P(\varphi)
\propto \exp\left[-N \tr V(\varphi)\right]$. While difficult to
evaluate diagrammatically, quite remarkably, this model can
be solved {\it for any} $V$, by using the orthogonal
polynomial approach\cite{bz1}. It is an intriguing
possibility that some analogous approach may be
developed for the type of problems discussed here, and for
large $N$ quantum chromodynamics (QCD). Another
approach towards solving large $N$ QCD involves the
concept of a master field. In the present context, this
amounts to finding a master field $h_{\rm master}$ which
dominates the functional integral in Eq.(\ref{field}).
Certainly, the KPZ problem is considerably simpler than
QCD, since we have one master field instead of the four
associated with $A_{\mu}$, $\mu=t, x,y,z$. However, we
do not have gauge invariance, which is powerful enough to
constrain the master fields in QCD to be independent of
space-time coordinates. Yet another approach is to consider
an RG flow \cite{bz2} in $N$, attempting to relate the theory
for $N+1$ to the one for $N$. The hope is that the model
will flow towards a fixed point as $N \rightarrow \infty$.
In any case, the field of large $N$ matrix theories is
evolving rapidly with recent advances in pure
mathematics\cite{v,fields}. Some of these advances may
ultimately prove fruitful in our context.

The calculations in this paper were limited to one loop
perturbation theory. Surprisingly, in the case of the
KPZ equation, distinct flow equations for the effective
coupling constant were obtained from two different
starting points (the dynamic equations, and an effective
field theory). However, the two RG schemes are
equivalent at the critical dimension, and appear to
result in the same exponents. We did not carry out
a similar program for the NS equation, as simple scaling
arguments appeared to be sufficient for obtaining
exponents. It would, nonetheless, be interesting to
check if a direct analysis reveals any surprises in this
case. By the same procedure as in Sec.\ref{sKPZ},
we can turn Eq.(\ref{matrixNS}) into a field theory.
The complicating new feature is the presence of
the new interaction term proportional to $\kappa$.
We don't know of any symmetry condition that will keep
$\kappa=0$ under renormalization. We also note that
the interaction terms, while still quartic in the velocity
fields, now contain only two derivatives, rather than
the four in the KPZ case.  It will interesting to study
the RG flows in the $(\lambda,\kappa)$ parameter
space. The previous scaling analysis will be invalid
at fixed points where either parameter is infinite (or
both are zero).

Finally, we conclude by noting yet another mapping of the
KPZ equation; to directed polymers in random
media\cite{MKlh}. For a {\it scalar field} $h({\bf x},t)$,
the transformation $U({\bf x},t)=\exp\left[
gh({\bf x},t)/2\right]$, changes the action in
Eq.(\ref{action2}) to
\begin{equation}\label{u}
S(U) = {2N\over g^2}\int d^d{\bf x}\, dt \, \tr \left(U^{-
1}\partial_t U - U^{-1}\partial^2_x U\right)^2.
\end{equation}
Promoting $U$ to an $N\times N$ matrix leads to another
possible generalization. We note that this is not, strictly
speaking, equivalent to Eq.(\ref{action2}) due to the
non-commutativity between matrices. However, we
can take this expression as another generalization,
no less valid than Eq.(\ref{action2}), of the KPZ
equation. The above action resembles a
non-relativistic version of non-linear sigma models, and
principal chiral field models, discussed in the field
theory literature, with the crucial difference that the
matrix $U$ is not unitary. Perhaps some of the
methods developed in this area can also be brought
to bear on the problem at hand.

\acknowledgements
One of us (AZ) thanks E. Br\'ezin for a discussion on coupled integral
equations in field theory. This work was supported in part by the National
Science
Foundation through Grant Nos. DMR-93-03667 (at MIT),
and PHY-89-04035 (at the ITP).

\begin{figure}
%\narrowtext
\caption{
The elements of a diagrammatic expansion starting with the
dynamical equation.
}
\label{F1}
\end{figure}

\begin{figure}
%\narrowtext
\caption{
The elements of a diagrammatic expansion starting with the
field theory action.
}
\label{F2}
\end{figure}

\begin{figure}
%\narrowtext
\caption{
The diagrams contributing to the one loop correction of the
propagator in the action.
}
\label{F3}
\end{figure}

%\end{multicols}

\begin{references}

\bibitem{KPZ}
M. Kardar, G. Parisi, and Y.-C. Zhang, Phys. Rev. Lett. {\bf
56}, 889 (1986).

\bibitem{reviews}
For recent reviews, see, e.g., {\it Dynamics of Fractal
Surfaces}, edited
by F. Family and T. Vicsek (World Scientific, Singapore,
1991); J. Krug and
H. Spohn, in {\it Solids Far From Equilibrium: Growth,
Morphology and
Defects}, edited by
C. Godr\`eche (Cambridge University Press, Cambridge,
1991); T.
Halpin--Healy and Y.-C. Zhang, Phys. Rep. {\bf 254}, 215
(1995); A. L.
Barabasi and H. E. Stanley, {\it Fractal concepts in surface
growth}
(Cambridge University press, Cambridge, 1995).

\bibitem{FNS}
D. Forster, D.R. Nelson, M. Stephen, Phys. Rev. A {\bf 16},
732 (1977).

\bibitem{Ala-Nissila}
T. Ala-Nissila et al., J. Stat. Phys. {\bf 72}, 207 (1993).

\bibitem{Medina}
E. Medina, T. Hwa, M. Kardar, and Y.-C. Zhang, Phys. Rev. {\bf
A39}, 3053
(1989).

\bibitem{Sun}
T. Sun and M. Plischke, Phys. Rev. E {\bf 49}, 5046 (1994).

\bibitem{Frey}
E. Frey and U. C. T\"auber, Phys. Rev. E {\bf 50}, 1024
(1994).

\bibitem{Schwartz}
M. Schwartz and S. F. Edwards, Europhys. Lett. {\bf 20}, 301
(1992).

\bibitem{Bouchaud}
J.-P.  Bouchaud and M. Cates, Phys. Rev. E {\bf 47}, R1455
(1993); Phys. Rev. E
{\bf 48}, 635 (1993).

\bibitem{Doherty}
J. P. Doherty, M. A. Moore, J. M. Kim, and A. J. Bray, Phys. Rev.
Lett.
{\bf 72}, 2041 (1994).

\bibitem{Tu}
Y. Tu, Phys. Rev. Lett. {\bf 73}, 3109 (1994).


\bibitem{Claudin}
M. A. Moore, T. Blum, J. P. Doherty, M. Marsili, J.-P. Bouchaud,
P.
Claudin, Phys. Rev. Lett. {\bf 74}, 4257 (1995).

\bibitem{Kolmogorov}
A. N. Kolmogorov, C. R. Acad. Sci. USSR {\bf 30}, 301 (1941);
{\it ibid.}
{\bf 32}, 16 (1941).

\bibitem{deDM}
C. DeDominicis and P. C. Martin, Phys. Rev. A {\bf 19}, 419
(1979).

\bibitem{YO}
V. Yakhot and S. A. Orszag, J. Sci. Comput. {\bf 1}, 3 (1986).

\bibitem{MW}
C.-Y. Mou and P. B. Weichman, Phys. Rev. Lett. {\bf 70},
1101 (1993); and
preprint (1994).

\bibitem{WIG}
E. Wigner, {\sl Can.\ Math.\ Congr.\ Proc.\/} p.174
(University of Toronto
Press); and other papers reprinted in Porter, op. cit.

\bibitem{POR}
C. E. Porter, {\it Statistical\ Theories\ of \ Spectra:\ \
Fluctuations\/}
(Academic Press, New York, 1965).

\bibitem{MEH}
M.L. Mehta, {\it Random\ Matrices\/}
(Academic Press, New York, 1991).

\bibitem{thooft}
G.~'t Hooft, Nucl. Phys. {\bf B72}, 461 (1974).

\bibitem{jerusalem}
See, for example, {\it Two Dimensional Quantum Gravity and
Random
Surfaces}, edited by D. Gross and T. Piran, (World Scientific,
Singapore,
1992).

\bibitem{bz}
See, for example, E. Br\'ezin and A. Zee, Phys. Rev. {\bf E49},
2588 (1994).

\bibitem{FDT}
U. Deker and F. Haake, Phys. Rev. A {\bf 11}, 2043 (1975).

\bibitem{ITU}
M. Kardar, Tr. J. of Physics {\bf 18}, 221 (1994).

\bibitem{Barma}
M. Barma and D. Dhar, Phys. Rev. Lett. {\bf 73}, 2135
(1994).

\bibitem{Ito}
J. Zinn--Justin, {\it Quantum Field Theory and Critical
Phenomena},
(Clarendon Press, Oxford, 1989), chapter 16.

\bibitem{noteN}
Note that while the perturbative corrections are valid for all
$N$, the RG
flows and exponents are only valid for $N=1$ and
$N\to\infty$. This is
because at other values of $N$ we must keep track of all
8 parameters
in Eq.(\ref{nKPZgeneral}).

\bibitem{Lassig}
M. Lassig, preprint cond-mat/9501094 (1995).

\bibitem{bz1} E. Br\'ezin and A. Zee, Nucl. Phys. {\bf 402}(FS),
613, (1993).

\bibitem{bz2}
E. Br\'ezin and A. Zee, Comp. Rend. Acad. Sci. (Paris) {\bf 317},
735 (1993).

\bibitem{v}
D. Voiculescu, K. Dykema, and A. Nica, {\it Free Random
Variables},
(American Mathematical Society, Providence, 1992).

\bibitem{fields} Lectures at the workshop on ``Operator
Algebra Free
Products and Random Matrices," Fields Insitute, March
1995, to appear in
the Proceedings.





\bibitem{MKlh}
M. Kardar, {\it Lectures on Directed Paths in Random Media},
Les Houches
Summer School on Fluctuating Geometries in Statistical
Mechanics and Field
Theory, August 1994 (to be published; see cond-
mat/9411022).


\end{references}
\end{document}